\documentclass[12pt]{article}
\usepackage{amsfonts,amsmath}
\usepackage{amssymb}

\usepackage{amsfonts}
\usepackage{dsfont}

\newcommand {\RN} [1] {\uppercase \expandafter {\romannumeral #1}}

\makeatletter \@addtoreset{equation}{section} \makeatother
\newcommand{\dr}{{{\rm d}}}

\usepackage[usenames]{color}
\usepackage{colortbl}
\usepackage{graphicx}
\usepackage{dcolumn}
\usepackage{bm}
\usepackage{amsmath}
\begin{document}
\begin{flushright}
FIAN/TD/07-22\\
\end{flushright}

\vspace{0.5cm}
\begin{center}
{\large\bf Disentanglement of
 Topological and Dynamical Fields in 3d Higher-Spin Theory within Shifted Homotopy Approach}

\vspace{1 cm}

\textbf{A.V.~Korybut${}^1$, A.A.~Sevostyanova${}^{1,2}$, M.A.~Vasiliev${}^{1,2}$ and V.A.~Vereitin${}^{1,2}$}\\

\vspace{1 cm}

\vspace{0.5 cm}
		\textbf{}\textbf{}\\
		\vspace{0.5cm}
		\textit{${}^1$ I.E. Tamm Department of Theoretical Physics,
			Lebedev Physical Institute,}\\
		\textit{ Leninsky prospect 53, 119991, Moscow, Russia}\\
		
		\vspace{0.7 cm} \textit{
			${}^2$ Moscow Institute of Physics and Technology,\\
			Institutsky lane 9, 141700, Dolgoprudny, Moscow region, Russia
		}

\par\end{center}

\begin{center}
\vspace{0.6cm}

\par\end{center}

\vspace{0.4cm}

\begin{abstract}
\noindent The first-order correction  to the one-form sector of equations   of the $3d$ higher-spin theory is derived from the generating nonlinear HS system by virtue of the shifted homotopy approach. The family of solutions to the generating system that disentangles equations for dynamical and topological fields in the first order of perturbation theory is found. This family is shown to belong to the different cohomology class compared to the solution found earlier by the direct methods.
The related cohomology  is shown to be the same as that underlying the mass deformation in the matter sector of $3d$ higher-spin equations.
\end{abstract}

\newpage
\tableofcontents

\newpage

\section{Introduction}

Full nonlinear higher-spin (HS) field theory in three dimensions is known at the level   of classical equations of motion which can be obtained from the generating system of \cite{Prokushkin:1998bq} (see also \cite{Vasiliev:1992ix,Vasiliev:1995dn}). HS vertices derived from the generating system can be written in the form
\begin{equation}\label{1form}
\dr \omega +\omega\ast \omega=\Upsilon(\omega,\omega,C)+\Upsilon(\omega,\omega,C,C)+\ldots,
\end{equation}
\begin{equation}\label{0form}
\dr C+\omega\ast C-C\ast \omega=\Upsilon(\omega,C,C)+\Upsilon(\omega,C,C,C)+\ldots\,,
\end{equation}
where the roles of the one-form $\omega$ and zero-form $C$ are specified below ($\dr:=dx^n\frac{\partial}{\partial x^n}$ is space-time de Rham derivative).

It is well known that $3d$ massless HS gauge fields do not propagate. However, in the sector of zero-forms $C$, the theory contains propagating {\it dynamical} matter fields.
These make the theory non-trivial even locally. In addition, the  zero-forms $C$ contain {\it topological} fields that do not propagate in the usual sense. These fields belong to the finite-dimensional modules over the isometry algera of $AdS_3$ which is a maximally symmetric solution of the $3d$ HS theory.  Topological fields do appear on the r.h.s. of \eqref{1form} and \eqref{0form} and the two types of fields may source each other.

For the analysis of the theory, it is desirable to disentangle the equations for dynamical and topological fields.
This problem was originally raised in \cite{Vasiliev:1992ix} where it has been solved in the first order of perturbation theory via particular field redefinition (this problem was also discussed in \cite{Kessel:2015kna}). In this paper we analyze the same problem in a more general setup appropriate for studying higher-order corrections in \eqref{1form}, \eqref{0form}. From this perspective, we are not interested in a particular field redefinition that solves the  problem in the first order, but examine a family of them resulting from the so-called {\it shifted homotopy approach} \cite{Gelfond:2018vmi, Didenko:2018fgx}.

Solving the equations of generating system \cite{Prokushkin:1998bq} one systematically faces  equations
\begin{equation}\label{df}
  \dr_z f=g\,,\qquad \dr_z:=dz^\alpha\frac{\partial}{\partial z^\alpha}
\end{equation}
on differential forms $f$ in certain auxiliary variables $z^\alpha$.
These equations can be solved with the help of the homotopy trick.
One of the advantages of the shifted homotopy approach of \cite{Didenko:2018fgx}
  is that, at least in the lowest orders of perturbation theory, it allows one to obtain local vertices with no additional field redefinitions.

The formalism originally proposed in \cite{Didenko:2018fgx} to the $4d$ HS theory has general applicability and can be applied to the HS theory in $AdS_3$.
The purpose of its application in this paper is  to solve equations of the generating system in such a way that the r.h.s. of  \eqref{1form} and \eqref{0form} for topological (dynamical) $\omega$ and $C$  be composed only of topological (dynamical) $\omega$ and $C$, thus disentangling   dynamical and topological fields. (Note that the locality problem, that emerges  at the second order, will be considered elsewhere.) Though the disentanglement problem  is successfully solved in this paper, the class of considered shifted homotopy operators  does not cover all  field redefinitions that  solve it in the first order. In particular, it does not reproduce the change of variables of \cite{Vasiliev:1992ix}.

An important observation of this paper, is on the role of the algebra of deformed oscillators \cite{Vasiliev:1989re} in the disentanglement  problem. Namely, it will be shown that the terms that govern the deformation of the $3d$ HS theory associated with the deformed oscillator algebra reproduce the cohomological terms that will be identified in the homotopy analysis.
Note that this observation may have another useful application in the context of the approach to holography relying on the unfolded formulation \cite{Vasiliev:2012vf}. It works as follows: one starts with equations for the free fields in $AdS$ and then by a proper rescaling\footnote{Two pairs of oscillators $y$ and $\bar{y}$ with canonical commutation relations \cite{Vasiliev:2012vf} get stretched in $\sqrt{z}$ times, where $z$ is the radial Puancar\'e coordinate in $AdS$.} of the oscillator variables goes to the boundary resulting in the equations on the
boundary conformal currents. The counterpart of this version of holography has  not yet been elaborated for massive fields in $AdS_3$.
Partially this is due to the highly involved product law of the deformed oscillator algebra \cite{Pope:1989sr,Bieliavsky:2008mv,Korybut:2014jza,Korybut:2020vmm}. However, cohomological (massive) terms obtained in this paper in terms
of the undeformed Moyal star product  may significantly simplify this analysis.

The layout of the rest of the paper is as follows. In section 2 the geometry of space-time and the frame-like formalism are recalled. Section 3 sketches elements of the $3d$ HS theory, including the language of two-component spinors, Moyal star product and the disentanglement  problem of topological and dynamical fields. Section 4 recalls the setup of the nonlinear HS equations of \cite{Prokushkin:1998bq}.
Section 5 briefly describes the homotopy approach, that reproduces perturbative solutions of the equations of the theory in the zeroth and first order. In sections 6, 7 the homotopy shift disentangling the topological and dynamical fields  is  obtained by searching for a gauge transformation.  In section 8 the cohomological plot~\cite{Barabanshchikov:1996mc} is completed while  in section 9 the homotopy shift is found again, but in a different way. Brief conclusions are given in section 10.

\section{Space-time geometry}

In this paper the  space-time geometry  will be described in terms of the one-form connection $A$
\begin{equation}
A=h^{a} P_{a}+\omega^{a b} L_{a b}\,,
\end{equation}
which takes values in the space-time symmetry algebra. Here generators of generalized translations $P_{a}$ and Lorenz rotations $L_{ab}$ are identified with the vielbein one-form $h^{a}=h_{\mu}^{a}(x) d x^{\mu}$ and spin-connection $\omega ^{a b}=-\omega ^{b a}=\omega _{\mu}^{ab}(x) dx^{\mu}$ with the convention $\mu,\nu =0,\ldots , d-1$, $a,b,\ldots , d-1$ (see \cite{MacDowell:1977jt}, \cite{Stelle:1979aj}).

The symmetry algebra of  $AdS_d$ is $o(d-1,2)$ with the gauge fields $A_{\mu}^{B C}=-A_{\mu}^{C B}$ (the indices $B, C=0, \ldots, d$ are raised and lowered by the flat metrics $\eta^{B C}=\operatorname{diag}(+-$ $\dots-+)$).
In terms of the connections $\omega_{\mu}^{a b}=A_{\mu}^{a b}, h_{\mu}^{a}=(\sqrt{2} \lambda)^{-1} A_{\mu}^{a d}$, the respective $o(d-1,2)$ gauge curvatures $\dr A+AA=L_{ab} R^{ab}+P_a R^a$ read as:
\begin{equation} \label{Curv}
R_{\mu \nu}{ }^{a b}=\partial_{\mu} \omega_{\nu}{ }^{a b}+\omega_{\mu}{ }^{a}{}_c \omega_{\nu}{ }^{c b}-2 \lambda^{2} h_{\mu}{ }^{a} h_{\nu}{ }^{b}-(\mu \leftrightarrow \nu),
\end{equation}
\begin{equation} \label{Tors}
R_{\mu \nu}{ }^{a}=\partial_{\mu} h_{\nu}{ }^{a}+\omega_{\mu}{ }^{a}{ }_{c} h_{\nu}{ }^{c}-(\mu \leftrightarrow \nu).
\end{equation}
Expression \eqref{Curv} differs from the Puancar\'e curvatures by the terms proportional to $\lambda^2$. One can express Lorentz connection $\omega_\mu{ }^{a b}$ via vielbein $h_\mu{ }^a$ with the aid of the constraint $R_{\mu \nu}{ }^a=0$ (for the non-degenerate $\left.h_\mu{ }^a\right)$. Substituting $\omega_\mu{ }^{a b}=\omega_\mu{ }^{a b}(h)$ into \eqref{Curv}, one can see that the condition $R_{\mu \nu}{ }^{a b}=0$ is equivalent to
\begin{equation}
\mathcal{R}_{\mu \nu}{ }^{a b}=2 \lambda^{2}\left(h_{\mu}{ }^{a} h_{\nu}{ }^{b}-h_{\nu}{ }^{a} h_{\mu}{ }^{b}\right),
\end{equation}
where $\mathcal{R}_{\mu \nu}{ }^{a b} := \partial_\mu \omega_\nu{ }^{a b}+\omega_\mu{ }^a{}_c \omega_\nu{ }^{c b}-(\mu \leftrightarrow \nu)$ is the Riemann tensor. Therefore it describes the AdS space-time with radius $(\sqrt{2} \lambda)^{-1}$:
\begin{equation}
\mathcal{R} := \mathcal{R}_{\nu \mu}{}^{\mu \nu}=-2  \lambda^{2} d(d-1).
\end{equation}

In the case of $AdS_3$ with $o(2,2) \sim sp(2) \oplus sp(2)$ it is convenient to use the spinor formalism where:
\begin{equation}
\omega^{\alpha \beta}(x)=\omega_{\nu}^{\alpha \beta}(x) d x^{\nu}, \quad h^{\alpha \beta}(x)=h_{\nu}^{\alpha \beta}(x) d x^{\nu}.
\end{equation}
Here $x^{\nu}$ are space-time coordinates $(\nu=0,1,2)$ and $\alpha, \beta=1,2$ are spinor indices which are raised and lowered with the aid of the symplectic form $\epsilon$:
\begin{equation}\label{index_rule}
\begin{array}{ll}
A_{\beta}=A^{\alpha} \epsilon_{\alpha \beta}, & A^{\alpha}=\epsilon^{\alpha \beta} A_{\beta}, \\
\epsilon_{\alpha \beta}=-\epsilon_{\beta \alpha}, & \epsilon_{\alpha \beta} \epsilon^{\alpha \gamma}=\delta_{\beta}^{\gamma} .
\end{array}
\end{equation}
The $AdS_3$ geometry is described by the zero-curvature and zero-torsion conditions
\begin{equation}\label{curv_spinor}
\dr \omega^{\alpha \beta}+\omega^{\alpha}{}_{\gamma} \wedge \omega^{\beta \gamma}+\lambda^{2} h^{\alpha}{}_{\gamma} \wedge h^{\beta \gamma}=0,
\end{equation}
\begin{equation}\label{tors_spinor}
\dr h^{\alpha \beta}+\omega^{\alpha}{}_{\gamma} \wedge h^{\beta \gamma}+h^{\alpha}{ }_{\gamma} \wedge \omega^{\beta \gamma}=0.
\end{equation}


\section{Fields and star product}

Such a formulation of space-time geometry provides a natural starting point for the formulation of the dynamics of free matter fields in terms of covariant constancy conditions for appropriate representations of the space-time symmetry algebra. Namely, this “unfolded formulation” \cite{Vasiliev:1992gr} allows one to rewrite free field equations in the form:
\begin{equation}
\dr C_{i}=A_{i}^{j} \wedge C_{j},
\end{equation}
where $\dr=d x^{\nu} \frac{\partial}{\partial x^{\nu}}$ is the space-time exterior differential and the gauge fields $A_{i}{ }^{j}=A^{a}\left(T_{a}\right)_{i}{ }^{j}$ obey the zero-curvature conditions:
\begin{equation}
\dr A^{a}=U_{b c}^{a} A^{b} \wedge A^{c},
\end{equation}
where $U_{b c}^{a}$ are structure coefficients of the space-time  Lie (super)algebra.

Next, it is convenient to represent the  scalar field zero-form in terms of the generating function \cite{Prokushkin:1998bq}
\begin{equation}
C\left(\hat{y} ; \psi_{1,2} ; k|x\right)=\sum_{A, B, C=0}^{1} \sum_{n=0}^{\infty} \frac{1}{n !} \lambda^{-\left[\frac{n}{2}\right]} C_{\alpha_{1} \ldots \alpha_{n}}^{A B C}(x) k^{A} \psi_{1}^{B} \psi_{2}^{C} \hat{y}^{\alpha_{1}} \ldots \hat{y}^{\alpha_{n}}\,.
\end{equation}
Here $\hat{y}^{\alpha_i}$ are Weyl oscillators, $k$ is the exterior Klein operator and $\psi_{1,2}$ are Clifford elements obeying the following conditions:
\begin{equation}
\begin{gathered}
\left[\hat{y}_{\alpha}, \hat{y}_{\beta}\right]=2 i \epsilon_{\alpha \beta}, \quad k \hat{y}_{\alpha}=-\hat{y}_{\alpha} k, \quad k^{2}=1,
\\
\left\{\psi_{i}, \psi_{j}\right\}=2 \delta_{i j}, \quad\left[\psi_{i}, \hat{y}_{\alpha}\right]=0, \quad\left[\psi_{i}, k\right]=0,\quad [\psi_i,dx^\mu]=0.
\end{gathered}
\end{equation}
The coefficients $ C_{\alpha_{1} \ldots \alpha_{n}}^{A B C}(x)$ are symmetric in the two-component spinor indices $\alpha$, that corresponds to the Weyl ordering in the Weyl algebra of the oscillators $\hat y_\alpha$.

Following \cite{Prokushkin:1998bq}, $\psi_1$ is introduced to describe the doubling of $sp(2)$ in the $AdS_3$ algebra $o(2,2)$, while the role of $\psi_2$ will be clarified below. The $o(2,2)$ generators can be realized as
\begin{equation}
L_{\alpha \beta}:= T_{\alpha \beta}\,,\qquad P_{\alpha \beta}:=\psi_1 T_{\alpha \beta}\,,\qquad T_{\alpha \beta}:=
\frac{1}{4 i}\left\{\hat{y}_{\alpha}, \hat{y}_{\beta}\right\}
\end{equation}
with the $sl_2\sim sp(2)$ generators $T_{\alpha \beta}$ obeying relations
\begin{equation}
\left[T_{\alpha \beta}, T_{\gamma \delta}\right]=\epsilon_{\alpha \gamma} T_{\beta \delta}+\epsilon_{\beta \delta} T_{\alpha \gamma}+\epsilon_{\alpha \delta} T_{\beta \gamma}+\epsilon_{\beta \gamma} T_{\alpha \delta}\,,
\end{equation}
\begin{equation}
\left[T_{\alpha \beta}, \hat{y}_{\gamma}\right]=\epsilon_{\alpha \gamma} \hat{y}_{\beta}+\epsilon_{\beta \gamma} \hat{y}_{\alpha}\,.
\end{equation}
So, the $o(2,2)$ one-form  connection can be realized as:
\begin{equation}
W_{g r}=\omega+\lambda h \psi_{1}, \quad \omega=\frac{1}{8 i} \omega^{\alpha \beta}\left\{\hat{y}_{\alpha}, \hat{y}_{\beta}\right\}, \quad h=\frac{1}{8 i} h^{\alpha \beta}\left\{\hat{y}_{\alpha}, \hat{y}_{\beta}\right\}.
\end{equation}
Now the flatness conditions \eqref{curv_spinor}, \eqref{tors_spinor}, that describe  $AdS_3$ geometry, read
\begin{equation}
\dr W_{g r}+W_{g r} \wedge W_{g r}=0.
\end{equation}

\subsection{Star product}
To write down covariant constancy equations for the set of fields $C\left(\hat{y} ; \psi_{1,2} ; k\right)$ it is convenient to replace all Weyl-ordered elements of the oscillator algebra  by their Weyl symbols according to the rule
\begin{equation}
C_{\alpha_{1} \ldots \alpha_{n}}^{A B C} k^{A} \psi_{1}^{B} \psi_{2}^{C} \hat{y}^{\alpha_{1}} \ldots \hat{y}^{\alpha_{n}} \rightarrow C_{\left(\alpha_{1} \ldots \alpha_{n}\right)}^{A B C} k^{A} \psi_{1}^{B} \psi_{2}^{C} y^{\alpha_{1}} \ldots y^{\alpha_{n}}\,,
\end{equation}
where $y^\alpha$ are usual commuting variables.

The commutation relations of the Weyl algebra are reproduced by the star product\footnote{The integration measure is normalized so that $(1 * f)(y)=(f * 1)(y)=f(y)$.}
\begin{equation}\label{Moyal}
(f * g)(y)=\int d u d v f(y+u) g(y+v) e^{i u_{\alpha} v^{\alpha}}\,,
\end{equation}
where $(f * g)(y)$ is the Weyl symbol of $f (\hat y) g(\hat y)$.
The following useful relations can be easily checked:
\begin{equation}
\left[y_{\alpha}, y_{\beta}\right]_{*}=2 i \epsilon_{\alpha \beta}, \quad\left[y_{\alpha}, f\right]_{*}=2 i \frac{\partial f}{\partial y^{\alpha}},
\end{equation}
\begin{equation} \label{com_yy}
\left[y_{\alpha} y_{\beta}, f(y)\right]_{*}=2 i\left(y_{\alpha} \frac{\partial}{\partial y^{\beta}}+y_{\beta} \frac{\partial}{\partial y^{\alpha}}\right) f(y),
\end{equation}
\begin{equation} \label{anticom_yy}
\left\{y_{\alpha} y_{\beta}, f(y)\right\}_{*}=2\left(y_{\alpha} y_{\beta}-\frac{\partial}{\partial y^{\alpha}} \frac{\partial}{\partial y^{\beta}}\right) f(y).
\end{equation}

\subsection{Topological and dynamical fields}

In these terms, the Weyl symbols for the $AdS_3$ connection and scalar field read as:
\begin{equation} \label{W_gr_sym}
W_{g r}=\omega+\lambda h \psi_{1}, \quad \omega=\frac{1}{4 i} \omega^{\alpha \beta} y_\alpha y_\beta, \quad h=\frac{1}{4 i} h^{\alpha \beta} y_\alpha y_\beta,
\end{equation}
\begin{equation}\label{C_comp}
C\left(y ; \psi_{1,2} ; k\right)=\sum_{A, B, C=0}^{1} \sum_{n=0}^{\infty} \frac{1}{n !} \lambda^{-\left[\frac{n}{2}\right]} C_{\alpha_{1} \ldots \alpha_{n}}^{A B C} k^{A} \psi_{1}^{B} \psi_{2}^{C} y^{\alpha_{1}} \ldots y^{\alpha_{n}}.
\end{equation}
One can define the full covariant derivative in $AdS_3$
\begin{equation}
D_{0} P=\dr P+W_{g r} * P-(-1)^{p} P * W_{g r},
\end{equation}
for a degree-$p$ differential form $P$, and write down  the covariant constancy equation   for the zero-form $C\left(y ; \psi_{1,2} ; k\right)$:
\begin{equation}
\label{DC}
D_{0} C = 0.
\end{equation}

Expansion of $C\left(y ; \psi_{1,2} ; k\right)$ in powers of  $\psi_2$
\begin{equation}\label{C^dt}
C\left(y ; \psi_{1,2} ; k\right)=C^{top}\left(y ; \psi_{1} ; k\right)+C^{dyn}\left(y ; \psi_{1}; k\right) \psi_{2},
\end{equation}
yields two equations
\begin{equation}
\label{D_top}
D C^{t o p}=-\lambda \psi_{1}\left[h, C^{t o p}\right]_{*},
\end{equation}
\begin{equation} \label{D_dyn}
D C^{d y n}=-\lambda \psi_{1}\left\{h, C^{d y n}\right\}_{*}\,,
\end{equation}
where $D:=\dr+[\omega, \bullet]_{*}$ is the Lorentz-covariant derivative.
So,  $\psi_2$ induces the decomposition of  equations (\ref{DC}) into two independent subsystems describing topological and dynamical fields.

Indeed, using the form of the generating function $C^{dyn} \left(y ; \psi_1 ; k\right)$ and vielbein $h(y)$, by virtue of \eqref{anticom_yy} one observes that equation \eqref{D_dyn} amounts to an infinite chain of equations:
\begin{equation}\label{DC_unfolded}
D C^{dyn}_{\alpha(n)}=\frac{\psi_1}{2 i}\left[h^{\gamma \delta} C^{dyn}_{\gamma \delta \alpha(n)}-\lambda^{2} n(n-1) h_{\alpha \alpha} C^{dyn}_{\alpha(n-2)}\right].
\end{equation}
All indices are raised and lowered by the rules \eqref{index_rule}. Here, symmetric sets of indices are denoted as $\alpha(n)=(\alpha_1 \dots \alpha_n)$.

It is easy to see that for lower $n$ these equations reduce to the massless Klein-Gordon and Dirac equations in $AdS_3$
\begin{equation}
D^{\mu} D_{\mu} C=\frac{3}{2} \lambda^{2} C, \quad h^{\nu}{}_{\alpha}{}^{\beta} D_{\nu} C_{\beta}=0\,.
\end{equation}
Other equations relate higher components in $y$ with the higher space-time derivatives of the scalar and spinor fields. Thus, given values of the fields $C_{\alpha(2 k)}(x_0)$ and $C_{\alpha(2 k+1)}(x_0)$ at some point $ x_0$ determines the behavior of the fields $C(y|x)$  in some its neighborhood.. This set of values is, in a sense, an infinite set of initial data in the Cauchy problem. Thus, not surprisingly, the dynamical scalar and spinor fields have an infinite number of degrees of freedom.

Analogously, one can obtain that \eqref{D_top} decomposes  into an infinite set of independent subsystems
\begin{equation}
D C_{\alpha(n)}^{t o p}=4 i \lambda n h_{\alpha}{ }^{\beta} C_{\beta \alpha(n-1)}^{t o p}
\end{equation}
with different $n$.
The number of initial data to be specified to solve the Cauchy problem is finite for every subsystem, being equal to the number of  components $C^{top}(y|x)$ of a fixed homogeneity degree  in $y$. Topological fields have a finite number of degrees of freedom and do not contribute to the local dynamics of the system. In fact, as emphasized in \cite{Didenko:2015pjo}, they play a role of coupling constants in the theory.
However, it is convenient to unify dynamical and topological fields into a single generating function to write a system of nonlinear equations in a compact way.

\section{Nonlinear system}

Following \cite{Prokushkin:1998bq}, to formulate the full nonlinear system that possesses necessary gauge symmetries and reduces at the linearized level to the free system, we introduce additional variables $z$ and three types of the generating functions
$W\left (z ; y ; \psi _{1,2} ; k\right )$,
$B\left (z ; y ; \psi _{1,2} ; k\right )$, and
$S\left (z ; y ; \psi _{1,2} ; k\right )$. 
The fields $W\left (z ; y ; \psi _{1,2} ; k\right )$ and $B\left (z ; y ; \psi _{1,2} ; k\right )$ extend the generating functions for HS and topological gauge fields $\omega(y; \psi_{1,2}; k)$ and massless matter fields and topological fields $C(y; \psi_{1,2}; k)$, respectively.

Generating function $S\left(z ; y ; \psi_{1,2} ; k\right)$ is a one-form in $dz$, having the meaning of connection in the additional variables
\begin{equation}
S\left(d z ; z ; y ; \psi_{1,2} ; k\right)=d z^{\alpha} S_{\alpha}\left(z ; y ; \psi_{1,2} ; k\right).
\end{equation}
The extended set of variables obeys the commutation relations
\begin{equation}
\begin{gathered}
{\left[y_{\alpha}, y_{\beta}\right]=\left[z_{\alpha}, z_{\beta}\right]=\left[z_{\alpha}, y_{\beta}\right]=0}, \\
\left\{d x_{\mu}, d x_{\nu}\right\}=\left\{d z_{\alpha}, d z_{\beta}\right\}=\left\{d z_{\alpha}, d x_{\mu}\right\}=0 ,\\
\left\{k, y_{\alpha}\right\}=\left\{k, z_{\alpha}\right\}=\left\{k, d z_{\alpha}\right\}=0, \quad k^{2}=1, \\
{\left[\psi_{i}, y_{\alpha}\right]=\left[\psi_{i}, z_{\alpha}\right]=\left[\psi_{i}, k\right]=0, \quad\left\{\psi_{i}, \psi_{j}\right\}=2 \delta_{i j}}.
\end{gathered}
\end{equation}

Also, $z$-variables bring new exterior differential $\dr_z$
\begin{equation}
\dr=d x^{\mu} \frac{\partial}{\partial x^{\mu}}, \quad \dr_{z}=d z^{\alpha} \frac{\partial}{\partial z^{\alpha}}, \quad \dr \dr_{z}=-\dr_{z} \dr.
\end{equation}
Further, the extension of the definition of the  star  product, taking into account the new variables,\footnote{Here we also normalize the measure in the way that $(1\ast f)(z,y)=(f\ast 1)(z,y)=f(z,y)$.}
\begin{equation}
f(z, y) * g(z, y)=\int d u d v f(z+u, y+u) g(z-v, y+v) e^{i u_{\alpha} v^{\alpha}}\,,
\end{equation}
has the following properties:
\begin{equation}
\begin{gathered}
{\left[y_{\alpha}, y_{\beta}\right]_{*}=-\left[z_{\alpha}, z_{\beta}\right]_{*}=2 i \epsilon_{\alpha \beta}, \quad\left[z_{\alpha}, y_{\beta}\right]_{*}=0}, \\
{\left[y_{\alpha}, f(z,y)\right]_{*}=2 i \frac{\partial f(z,y)}{\partial y^{\alpha}}, \quad\left[z_{\alpha}, f(z,y)\right]_{*}=-2 i \frac{\partial f(z,y)}{\partial z^{\alpha}}}, \\
e^{i z_{\alpha} y^{\alpha}} * f(z, y)=f(-z,-y) * e^{i z_{\alpha} y^{\alpha}}.
\end{gathered}
\end{equation}
The exponent in the last equation $\varkappa = e^{i z_{\alpha} y^{\alpha}}$ is conventionally called the inner Klein operator.

In these terms one can write the zero-curvature and covariant constancy equations for the fields $W$, $B$ and $S$ to obtain a nonlinear system of equations of \cite{Prokushkin:1998bq} of the form
\begin{equation}\label{PV_1}
	\dr W + W * W = 0,
\end{equation}
\begin{equation}\label{PV_2}
	\dr B + W * B - B * W = 0,
\end{equation}
\begin{equation}\label{PV_3}
	\dr S + W * S + S * W = 0,
\end{equation}
\begin{equation}\label{PV_4}
	S * S = i(dz^{\alpha}dz_{\alpha}+B*dz^{\alpha}dz_{\alpha} e^{i z_\alpha y^\alpha} k),	
\end{equation}
\begin{equation}\label{PV_5}
	S * B - B * S = 0.
\end{equation}

\section{Shifted homotopy}

\subsection{Homotopy trick}

Solving the nonlinear HS system, one faces equations of the type
\begin{equation} \label{dz_f}
\dr_{z} f(z ; y ; d z)=J(z ; y ; d z), \text{ where } \dr_{z} J(z ; y ; d z)=0\,,\quad J(z ; y ; 0)=0.
\end{equation}
In the HS context, such equations were originally analyzed in the $4d$ theory  \cite{Vasiliev:1992av} by the homotopy technique in the particular case of the so-called conventional homotopy.
In \cite{Didenko:2018fgx} this approach was further generalized to  shifted homotopies proven to be useful to analyze the locality problem in $4d$ HS theory.
Namely, a particular solution of \eqref{dz_f} can be found in the form
\begin{equation}
f_q(J)=\Delta_{q} J(z ; y ; d z):=\left(z^{\alpha}+q^{\alpha}\right) \frac{\partial}{\partial d z^{\alpha}} \int_{0}^{1}  \frac{d t}{t} J(t z-(1-t) q ; y ; t d z)\, .
\end{equation}
Here $q$ is $z$- and $dz$-independent spinor which in general can be an operator. The general solution of \eqref{dz_f} is
\begin{equation}
f(z ; y ; d z)=f_{q}(J)(z ; y ; d z)+h(y;dz)+\dr_{z} \epsilon(z ; y ; d z),
\end{equation}
where $h(y;dz)$ is in $\dr_z$-cohomology and $\dr_{z} \epsilon(z ; y ; d z)$ is an exact form. At $q_\alpha\neq 0$ homotopy is called {\it shifted} as it results from the shift of $z_\alpha$ by $q_\alpha$.

The action of two operators can be written in a symmetric form as an integral over a triangle in a three-dimensional space:
\begin{equation}
\begin{split}
\Delta_{p} \Delta_{q} f(z ; y ; d z)=\int_{[0,1]^{3}} d^{3} \tau \delta\left(1-\tau_{1}-\tau_{2}-\tau_{3}\right) &\left(z^{\alpha}+p^{\alpha}\right)\left(z^{\beta}+q^{\beta}\right)
\times \\ \times &
\frac{1}{\tau_1^2} \frac{\partial^2}{\partial dz^\alpha \, \partial dz^\beta} f(\tau_{1} z-\tau_{3} p-\tau_{2} q ; y; \tau_1 dz)\,.
\end{split}
\end{equation}

Introducing the projection operator onto the cohomology space $h_{q} f(z ; d z)=f(-q ; 0)$, one obtains, following \cite{Didenko:2018fgx}, a list of useful relations
\begin{equation}
\Delta_{p} \Delta_{q}=-\Delta_{q} \Delta_{p}, \quad h_{p} \Delta_{q}=-h_{q} \Delta_{p}, \quad h_{p} \Delta_{p}=0, \quad
h_{p} h_{q}=h_{q}, \quad \Delta_{q} h_{p}=0,
\end{equation}
\begin{equation}\label{resId}
\left\{\dr_{z}, \Delta_{q}\right\}=1-h_{q},
\qquad
\Delta_{b}-\Delta_{a}=\left[\dr_{z}, \Delta_{a} \Delta_{b}\right]+h_{a} \Delta_{b}.
\end{equation}

It is particularly useful to compute the action of the  homotopy operators  on the central element $\gamma=dz^{\alpha}dz_{\alpha} e^{i z_\alpha y^\alpha} k$ from the $r.h.s$ of \eqref{PV_4}. Application of $h_c\Delta_b \Delta_a$ to $\gamma$ yields
\begin{equation} \label{hDD_g}
h_{c} \Delta_{b} \Delta_{a} \gamma=2 \int_{[0,1]^{3}} d^{3} \tau \delta\left(1-\tau_{1}-\tau_{2}-\tau_{3}\right)(c-b)_{\gamma}(c-a)^{\gamma} e^{-i\left(\tau_{1} c+\tau_{2} a+\tau_{3} b\right)_\alpha y^\alpha} k.
\end{equation}

Using the resolution identity \eqref{resId}  and the fact that $\gamma$ is a two-form in $dz$, one  obtains the following relations often used in the shifted homotopy computations
\begin{equation}
\left(\Delta_{b}-\Delta_{a}\right) \gamma=\dr_{z} \Delta_{a} \Delta_{b} \gamma,
\end{equation}
\begin{equation}\label{Tr}
\Delta_{c}\left(\Delta_{b}-\Delta_{a}\right) \gamma=\left(h_{c}-1\right) \Delta_{b} \Delta_{a} \gamma,
\end{equation}
\begin{equation}
\left(\Delta_{d}-\Delta_{c}\right)\left(\Delta_{b}-\Delta_{a}\right) \gamma=\left(h_{d}-h_{c}\right) \Delta_{b} \Delta_{a} \gamma,
\end{equation}
\begin{equation}
\left(\Delta_{c} \Delta_{b}-\Delta_{c} \Delta_{a}+\Delta_{b} \Delta_{a}\right) \gamma=h_{c} \Delta_{b} \Delta_{a} \gamma.
\end{equation}

\subsection{Star-exchange formulas}

Since equations \eqref{PV_1}-\eqref{PV_5} are written in the star-product formalism, it is important to understand how the shifted homotopy operators interact with the star product. Here we present some of the relevant relations called star-exchange formulas originally obtained in \cite{Didenko:2018fgx}.

Let the homotopy operator $\Delta_{q+\alpha y}$ act on the star product $A(y;k) * \phi(z;y;k;dz)$, where $A(y;k)$ is a space-time $r$-form, $q$ is $y$-independent  and $\alpha$ is a $\mathds{C}$-number. Then the relation
\begin{equation}\label{star_(1-a)p}
	\Delta_{q+\alpha y} (A(y;k) * \phi(z;y;k;dz)) = (-1)^r A(y;k) * \Delta_{q+(1-\alpha)p+\alpha y} \phi(z;y;k;dz)
\end{equation}
holds true.
Here  a spinor differential operator $p_\alpha$ is defined to  act according to the rule
\begin{equation} \label{def_p}
	p_\alpha A(y;k) \equiv A(y;k) p_\alpha := -i \dfrac{\partial}{\partial y^\alpha} (A_1(y)+A_2(y)k).
\end{equation}

Note that the operator $p_\alpha$ differentiates $A(y; k)$ with respect to its full argument and commutes with all other symbols. Let us stress that the operator $p^\alpha$ with upper spinor index has  opposite sign by definition
\begin{equation}
	p^\alpha A(y;k) := +i \dfrac{\partial}{\partial y_\alpha} A(y;k),
	\qquad
	p^\alpha = \epsilon^{\alpha \beta} p_\beta.
\end{equation}
Note that in these terms the  star product of two $y$-functions can be written as
\begin{equation}
	F(y) * G(y) = F(y-p_G)G(y)= F(y)G(y+p_F).
\end{equation}
For central element $\gamma=dz^{\alpha}dz_{\alpha} e^{i z_\alpha y^\alpha} k$ by virtue of star-exchange relation \eqref{star_(1-a)p} one can show that
\begin{equation}\label{star_2p}
\Delta_{\tilde{q}} \gamma * A(y ; k)= (-1)^r A(y ; k) * \Delta_{\tilde{q}+2 p} \gamma, \quad \tilde{q}=q+\alpha y\, .
\end{equation}

\section{Perturbative expansion}

\subsection{Vacuum solution}

We will analyze  the system \eqref{PV_1}-\eqref{PV_5} perturbatively. Let us start with the vacuum solution with $B = B_0 = 0, \: W = W_0, \: S = S_0$. The system takes the form
\begin{equation}\label{PV_1.0}
	\dr W_0 + W_0 * W_0 = 0,
\end{equation}
\begin{equation}\label{PV_3.0}
	\dr S_0 + W_0 * S_0 + S_0 * W_0 = 0,
\end{equation}
\begin{equation}\label{PV_5.0}
	S_0 * S_0 = i dz^{\alpha}dz_{\alpha}.
\end{equation}
Equation \eqref{PV_1.0} is solved by $W_0=W_{gr}$ that describes the $AdS_3$ background connection
\begin{equation}\label{W_0}
	W_0(y|x) = \omega_0(y|x) + \lambda h_0(y|x) \psi_1,
	\quad
	\omega_0(y|x) = \dfrac{1}{4i} \omega_0^{\alpha \beta}(x) y_\alpha y_\beta\,,
	\quad
	h_0(y|x) = \dfrac{1}{4i} h_0^{\alpha \beta}(x) y_\alpha y_\beta\,.
\end{equation}

Equations \eqref{PV_3.0}, \eqref{PV_5.0} are solved by $S_0 = dz^\alpha z_\alpha$. Note that the star-commutator with $S_0$ therefore acquires the meaning of the $z_\alpha$ exterior derivative
\begin{equation}\label{S0_dz}
	[S_0,f]_* = -2i dz^\alpha \dfrac{\partial f}{\partial z^\alpha} = -2i \dr_z f.
\end{equation}


\subsection{Linearisation}

In the first order of the perturbative expansion, we put $B=C$, $W=W_0+W_1$, $S=S_0+S_1$. Then  equations \eqref{PV_1}--\eqref{PV_5} yield
\begin{equation}\label{PV_1.1}
\dr W_{1}+W_{0} * W_{1}+W_{1} * W_{0}=0,
\end{equation}
\begin{equation}\label{PV_2.1}
\dr C+W_{0} * C-C * W_{0}=0,
\end{equation}
\begin{equation}\label{PV_3.1}
\dr S_{1}+W_{0} * S_{1}+S_{1} * W_{0}=-\left\{S_{0}, W_{1}\right\}_{*},
\end{equation}
\begin{equation}\label{PV_4.1}
\left\{S_{0}, S_{1}\right\}_{*}=i C * \gamma,
\end{equation}
\begin{equation}\label{PV_5.1}
\left[S_{0}, C\right]_{*}=0.
\end{equation}
From \eqref{PV_5.1} and \eqref{S0_dz} one can see that $C$ is indeed $z$-independent,  $C=C(y;\psi_{1,2};k|x)$. From \eqref{PV_4.1} we get
\begin{equation} \label{dz_S1}
\dr_z S_1 (z; dz; y; \psi_{1,2}; k|x) = - \frac{1}{2} C(y; \psi_{1,2}; k|x) * \gamma.
\end{equation}

Generally, the field $C(y;\psi_{1,2};k|x)$ is the sum of the $k$-independent part and that linear in $k$.
For the sake of simplicity, below we consider only the part linear in $k$ and, slightly abusing the notation, we denote it $C(y;\psi_{1,2}|x)k$. For the $k$-independent part all the computations are analogous, and for even background connections $W_0(-y)=W_0(y)$ they are identical. Since  $k$ anticommutes with $y$, we will use notation $k A(y) = A(-y) k := \tilde{A}(y) k$.

Applying the homotopy trick with the homotopy parameter $q=0$ and using star-exchange formulas, one obtains
\begin{equation} \label{10}
S_1(z; dz; y; \psi_{1,2}; k|x) = -\frac{1}{2} C(y; \psi_{1,2}|x) * \Delta_p \gamma k.
\end{equation}
Plugging the expression for $S_1$ into \eqref{PV_3.1} and using \eqref{S0_dz}, one finds, keeping $\psi_{1,2}$ and $x$ implicit
\begin{equation} \label{dz_W}
	\dr_z W_1 = - \frac{1}{4i} \left[ W_0 * C * (\Delta_{p}-\Delta_{p+t})\gamma - C * \tilde{W}_0 * (\Delta_{p-2t}-\Delta_{p-t})\gamma \right]k := J_W(z;y;dz).
\end{equation}
Solution to this equation  via conventional homotopy (with parameter $q=0$) yields
\begin{equation}\label{W_1_0}
	W_1(y;z) = \omega(y) + \Delta_0 J_W(y;z;dz) = \omega + W_1^0.
\end{equation}
The $z$-independent field $\omega(y)$ serves as the generating function for HS gauge fields corresponding to the pure Chern-Simons HS theory of Blencowe \cite{Blencowe:1988gj}. Dynamical and topological sectors in space-time one-forms are distinguished between in a way opposite to \eqref{C^dt},
\begin{equation}
	\omega (y ; \psi_{1,2})=\omega^{dyn}\left(y;\psi_{1}\right)+\omega^{top}\left(y;\psi_{1}\right) \psi_{2}\,,
\end{equation}
with the $\psi_2$-independent component identified as dynamical while that linear in $\psi_2$ as topological (such a complementary picture for one- and zero-forms also occurs in $4d$ HS theory \cite{Vasiliev:1988sa}). In particular, the gravitational fields \eqref{W_0} belong to the dynamical sector.

Substitution of  \eqref{W_1_0} into \eqref{PV_1.1} yields
\begin{equation}\label{D0_w}
    D_0 \omega = - D_0 W_1^0 = \Delta R (W_0,C).
\end{equation}

For the $4d$ problem, the analogous deformation $\Delta R$ is non-zero encoding non-trivial HS equations, the so-called central on-mass-shell theorem \cite{Vasiliev:1988sa}. Since $3d$ HS fields do not propagate, the deformation $\Delta R$ is anticipated to be trivial.
Analysing the dynamical and topological sectors it is easy to see that in the $\psi_2$-independent dynamical
sector indeed $\Delta R^{dyn} = 0$. However, in the topological sector
\begin{equation}\label{D_R}
    \Delta R^{top}\left(y ; \psi_{1}\right)=-\left.\frac{\lambda^2}{16i} h_{0}{ }^{\alpha}{}_{\gamma} \wedge h_{0}{}^{\gamma \beta}\left(y_{\alpha}-p_{\alpha}\right)\left(y_{\beta}-p_{\beta}\right) C^{dyn}\left(\xi ; \psi_{1}\right)\right|_{\xi=0}\neq 0\,.
\end{equation}

We observe that the deformation in the topological sector is expressed in terms of the dynamical fields.
Hence, in this setup topological fields will contribute to the HS field equations in higher orders. This mixing of dynamical and topological sectors complicates  further higher order calculations.
It turns out to be possible to compensate the deformation of the  $r.h.s$ of \eqref{D0_w} by an appropriate shift (field redefinition) of variables of the structure $\omega^{top} \rightarrow \omega^{top}+h_{0} C^{dyn}$ with some space-time local $C^{dyn}$-dependent terms. Such a field redefinition   originally  found in \cite{Vasiliev:1992ix}  is
\begin{equation}
    \omega^{top}(y)=\omega^{\prime \,top}(y)+\delta \omega^{top} (y),
\end{equation}
\begin{equation}\label{dw_old}
	\delta \omega^{top} (y)=-\frac{\lambda}{8i} \int_{0}^{1} d t\left(1-t^{2}\right) h_{0}^{\alpha \beta}\left(y_{\alpha}-p_{\alpha}\right)\left(y_{\beta}-p_{\beta}\right) C^{d y n}(t y) \psi_1 .
\end{equation}

Our goal is to find all solutions of equation \eqref{dz_W} within the  shifted homotopy approach, that trivialize the  $r.h.s.$ of \eqref{D0_w} in order to avoid mixing dynamical and topological fields.

\section{Shifted solution}\label{section7}

Our aim is to solve the equation for the master field $W_1$ in such a way that the first-order correction in $C$ to the sector of one-form equations is trivial, {\it i.e.}  $W_1$ should obey the equation
\begin{equation}\label{D0W1}
D_0(W_1)=0.
\end{equation}

\subsection{Shifted homotopy setup}\label{Shifted_homotopy_setup}

Equation  \eqref{dz_W}, that reconstructs the $z$-dependence of $W_1$, can schematically  be  put into the form
\begin{equation}\label{terms}
d_z W_1=(\ldots)^{top}+(\ldots)^{top}\psi_1+(\ldots)^{dyn}\psi_2+(\ldots)^{dyn}\psi_1 \psi_2 .
\end{equation}
Here superscripts $top$ and $dyn$ indicate if the expressions in the brackets are composed of $C^{top}(y;\psi_1)$ or $C^{dyn}(y;\psi_1)$.
Note that here the dependence of the $C$-fields on $\psi_1$ is implicit, while the explicit dependence on $\psi_1$ in \eqref{terms} comes through the background $AdS_3$ connection \eqref{W_0}. Since $\dr_z$ commutes with $\psi_{1,2}$, each expression in brackets in \eqref{terms} is $\dr_z$-closed. Hence, to obtain a solution, one can apply different homotopy operators to each of them.
For example, one can apply conventional homotopy to all of these terms to obtain a particular solution for $W_1=W_1^0$
\begin{equation}\label{W_1^0}
W_1^0 = \frac{1}{4 i} \left[W_{0}*C*\Delta_{p+t}\Delta_{p} \gamma - C * \tilde{W}_{0}*\Delta_{p-t}\Delta_{p-2 t} \gamma \right]k\, ,
\end{equation}
where the superscript $0$ indicates that the solution results from the zero-shift (conventional) homotopy  $\Delta_0$.
Straightforward computation shows that only the term proportional to $\psi_1 \psi_2$ (recall that additional dependence on $\psi_1$ is hidden in $C^{dyn}(y,\psi_1)$) of $W_1$ computed with the help of conventional homotopy is not covariantly constant: its covariant derivative is equal to $-\Delta R$  $\eqref{D_R}$. For that reason we  apply shifted homotopy only to the part of the {\it r.h.s.} of \eqref{dz_W}  proportional to $\psi_1 \psi_2$.
Also, we use the fact that the  system of \cite{Prokushkin:1998bq} is consistent  with all fields  valued in any associative algebra, which implies that terms with different orders of $C$ and $W_0$ are $\dr_z$-closed independently.
The part of the {\it r.h.s.} of \eqref{dz_W} we are interested in is
\begin{equation}
-\frac{\lambda}{4 i}\left[h_{0} * C^{dyn} *  \big(\Delta_{p+t}- \Delta_{p}\big) \gamma +C^{dyn} * \tilde{h}_0 * \big(\Delta_{p-2t}-\Delta_{p-t}\big) \gamma \right] \psi_1 \psi_2 k.
\end{equation}
We apply the homotopy operator $\Delta_{\mu_1 p+\nu_1 t+\alpha_1 y}$ to the first term and $\Delta_{\mu_2 p-\nu_2 t+\alpha_2 y}$ to the second. Using \eqref{Tr} one can check that the difference $\delta W_1$  between $W_1$ solved  this way and $W_1^0$ \eqref{W_1^0} is
\begin{equation}\label{dW_ddg}
\delta W_1 = -\frac{\lambda}{4 i}\left[h_{0} * C^{dyn} * h_{Q_{1}(p,t,y)} \Delta_{p+t} \Delta_{p} \gamma +C^{dyn} * \tilde{h}_0 * h_{Q_2(p,-t,y)} \Delta_{p-t} \Delta_{p-2t} \gamma \right] \psi_1 \psi_2 k,
\end{equation}
where
\begin{equation}\label{Qi}
	Q_i(p,t,y) = (\mu_i-\alpha_i+1) p  + (\nu_i-\alpha_i+1) t + \alpha_i y.
\end{equation}
Expressions \eqref{Qi} result from $\Delta_{\mu_1 p+\nu_1 t+\alpha_1 y}$ and $\Delta_{\mu_2 p-\nu_2 t+\alpha_2 y}$ via star-exchange relations \eqref{star_(1-a)p}, \eqref{star_2p}.
To arrive at \eqref{D0W1}, the parameters $\mu_{1,2}$, $\nu_{1,2}$, $\alpha_{1,2}$ have to be adjusted in such a  way that
\begin{equation}\label{Goal}
D_0 (\delta W_1)=\Delta R\,.
\end{equation}

Applying formula \eqref{hDD_g} and the Taylor expansion, one can write
\begin{multline} \label{dW_explicit}
\delta W_1 =- \frac{\lambda}{2i} \int_{[0,1]^3} d^3 \tau \delta(1-\tau_1-\tau_2-\tau_3) \times\\
\Big[(\mu_1 p+\alpha_1 y)_\gamma t^\gamma C^{dyn}\big(-(\mu_1-\alpha_1)\tau_1 y,\psi_1\big) h_0 \big( (\tau_2-(\nu_1-\alpha_1)\tau_1)y - (1-\tau_2+(\nu_1-\mu_1) \tau_1)p \big)+\\
+(\mu_2 p+\alpha_2 y)_\gamma t^\gamma C^{dyn}\big(-(\mu_2-\alpha_2)\tau_1 y,\psi_1\big) h_0 \big( (\tau_2+(\nu_2-\alpha_2)\tau_1)y -(1-\tau_2-(\nu_2-\mu_2) \tau_1)p \big)
    	\Big] \psi_1 \psi_2.
\end{multline}
Then, plugging in the $AdS_3$-vielbein $h_0(y) = \frac{\lambda}{4i} h_0{}^{\alpha \beta} y_\alpha y_\beta$ and integrating over  $\tau_2$ and $\tau_3$, one obtains
\begin{multline}\label{dW_int}
\delta W_1=-\frac{\lambda}{4i}\int_0^1 d t\, (1-t)\, h_0^{\alpha \beta} \times\\
\times \Big\{\alpha_1 \Big(\frac{1-t}{2}-(\nu_1-\alpha_1)t\Big)y_\alpha y_\beta C^{dyn}((\alpha_1-\mu_1)t y,\psi_1)+\\
+\alpha_2 \Big(\frac{1-t}{2}+(\nu_2-\alpha_2)t\Big)y_\alpha y_\beta C^{dyn}((\alpha_2-\mu_2)ty,\psi_1)+\\
+\Big[\mu_1\Big(\frac{1-t}{2}-(\nu_1-\alpha_1)t\Big)-\alpha_1\Big(1-\frac{1-t}{2}+(\nu_1-\mu_1)t\Big)\Big]y_\alpha p_\beta C^{dyn}((\alpha_1-\mu_1)ty,\psi_1)+\\
+\Big[\mu_2\Big(\frac{1-t}{2}+(\nu_2-\alpha_2)t\Big)-\alpha_2\Big(1-\frac{1-t}{2}-(\nu_2-\mu_2)t\Big)\Big]y_\alpha p_\beta C^{dyn}((\alpha_2-\mu_2)ty,\psi_1)-\\
-\mu_1\Big(1-\frac{1-t}{2}+(\nu_1-\mu_1)t\Big)p_\alpha p_\beta C^{dyn}((\alpha_1-\mu_1)ty,\psi_1)-\\
-\mu_2\Big(1-\frac{1-t}{2}-(\nu_2-\mu_2)t\Big)p_\alpha p_\beta C^{dyn}((\alpha_2-\mu_2)ty,\psi_1)\Big\}\psi_1 \psi_2.
\end{multline}

To solve the problem, one has to act by the derivative $D_0$ on this expression and choose the homotopy parameters in such a way that the result yields  the deformation $\Delta R$.

\subsection{Covariant derivative}

The expression  \eqref{dW_int}  is the sum of the one-forms of the types
\begin{subequations} \label{ABC}
\begin{equation}\label{A}
	\mathrm{A}= - \frac{\lambda}{4i} h_0^{\alpha \beta} \int_{0}^{1} d t\, a(t) y_{\alpha} y_{\beta} C^{d y n}(\rho t y,\psi_1) \psi_{1} \psi_{2},
\end{equation}
\begin{equation}\label{B}
	\mathrm{B}= - \frac{\lambda}{4i} h_0^{\alpha \beta} \int_{0}^{1} d t\, b(t) y_{\alpha} p_\beta C^{d y n}(\rho t y,\psi_1) \psi_{1} \psi_{2},
\end{equation}
\begin{equation}\label{C}
	\mathrm{C}= - \frac{\lambda}{4i} h_0^{\alpha \beta} \int_{0}^{1} d t\, c(t) p_\alpha p_\beta C^{d y n}(\rho t y,\psi_1) \psi_{1} \psi_{2},
\end{equation}
\end{subequations}
where $a(1)=b(1)=c(1)=0$ and $\rho$ denotes some coefficients that may differ for $A$, $B$ and $C$.
To evaluate the action of the covariant derivative on them, one has to use the Schouten identity and integration by parts
\begin{equation}
	h_0{}^{\alpha \beta} \wedge h_0{}^{\gamma \delta} = \frac{1}{2} H^{\alpha \gamma} \epsilon^{\beta \delta} + \frac{1}{2} H^{\beta \delta} \epsilon^{\alpha \gamma}, \quad \text{where} \quad
	H^{\alpha \beta} = 	h_0{}^{\alpha}{}_{\gamma} \wedge h_0{}^{\gamma \beta},
\end{equation}
\begin{equation}
	\int_0^1 dt\, a(t) y^\sigma \frac{\partial}{\partial (\rho t y^\sigma)} C^{dyn}(\rho t y,\psi_1) =  \frac{a(t)}{\rho} C^{dyn}(\rho t y,\psi_1) \bigg|_0^1 - \int_0^1 dt\, \frac{1}{\rho} \frac{\partial a(t)}{\partial t} C^{dyn}(\rho t y,\psi_1).
\end{equation}
As a result,  the action of the covariant derivative takes the form of the sum of the boundary term proportional to $C^{dyn}(0,\psi_1)$ and the integral containing a differential operator acting in the same way on the measures $a(t), b(t), c(t)$:
\begin{subequations}\label{D0_ABC}
\begin{align}
	&\begin{aligned}
	D_0 \mathrm{A} = \frac{\lambda^2}{8i} &\bigg[ \frac{a(0)}{\rho} H^{\alpha \beta} y_\alpha p_\beta C^{d y n}(\xi,\psi_1) \big|_{\xi=0}+
	\\
	&+ \int_0^1 dt\, \big( 2\rho t a(t)-\rho t^2 \frac{\partial a(t)}{\partial t}+\frac{1}{\rho} \frac{\partial a(t)}{\partial t} \big) H^{\alpha \beta} y_\alpha p_\beta C^{d y n}(\rho t y,\psi_1) \bigg],
	\end{aligned}
\\
	&\begin{aligned}
	D_0 \mathrm{B} = \frac{\lambda^2}{16i} &\bigg[ \frac{b(0)}{\rho} H^{\alpha \beta} [y_\alpha y_\beta + p_\alpha p_\beta] C^{d y n}(\xi,\psi_1) \big|_{\xi=0}+
	\\
	&+ \int_0^1 dt \,\big( 2\rho t b(t)-\rho t^2 \frac{\partial b(t)}{\partial t}+\frac{1}{\rho} \frac{\partial b(t)}{\partial t} \big) H^{\alpha \beta} [y_\alpha y_\beta + p_\alpha p_\beta] C^{d y n}(\rho t y,\psi_1) \bigg],
	\end{aligned}
\\
	&\begin{aligned}
	D_0 \mathrm{C} = \frac{\lambda^2}{8i} &\bigg[ \frac{c(0)}{\rho} H^{\alpha \beta} y_\alpha p_\beta C^{d y n}(\xi,\psi_1) \big|_{\xi=0}+
	\\
	&+ \int_0^1 dt\, \big( 2\rho t c(t)-\rho t^2 \frac{\partial c(t)}{\partial t}+\frac{1}{\rho} \frac{\partial c(t)}{\partial t} \big) H^{\alpha \beta} y_\alpha p_\beta C^{dyn}(\rho t y,\psi_1) \bigg].
	\end{aligned}
\end{align}
\end{subequations}

Note that $D_0 A$ and $D_0C$ have the same form. This is not incidental, being a consequence of the fact that $A-C$ is $D_0$-exact. This follows from the observation that, as is not hard to check,  for any zero-form  field $\epsilon=C^{dyn}(\rho t y) \psi_2$,
\begin{equation}\label{D0exact}
    D_0 \epsilon = -\frac{\lambda}{2i} (1-\rho^2 t^2) h_0^{\alpha \beta} (y_{\alpha}y_{\beta} - p_{\alpha} p_{\beta}) C^{dyn}(\rho t y,\psi_1) \psi_1 \psi_2\,.
\end{equation}
Since $D_0^2=0$, it follows that $D_0(A-C)=0$. Hence it is enough to consider $A+C$. For instance, one can set $C=0$.

For $D_0(A+B+C)$ to give $\Delta R$ the integrands have to vanish (only boundary terms appear in \eqref{D_R}). Solving the differential equation, one finds that the measures $a(t)$, $b(t)$, $c(t)$ in \eqref{ABC} must have the form:
\begin{equation}\label{diff_eq}
	f(t) = (1-\rho^2 t^2) \cdot \text{const}.
\end{equation}
Note that, if the polynomials are proportional to $(1-t)$ like in \eqref{dW_int}, this is possible only for $\rho=\pm 1$. Hence, expression \eqref{dW_int} should acquire the form
\begin{equation}
	\begin{gathered}
		\delta W_1 =\label{co}
		- \frac{\lambda}{4i} \int_0^1 dt\,  h_0{}^{\alpha \beta} \cdot
		\\
		\begin{aligned}
    		\cdot \big[
    		a_1(t) y_\alpha y_\beta C^{dyn}((\alpha_1-\mu_1)t y,\psi_1) &+
    		a_2(t) y_\alpha y_\beta C^{dyn}((\alpha_2-\mu_2)t y,\psi_1)
    		+\\+
    		b_1(t) y_\alpha p_\beta C^{dyn}((\alpha_1-\mu_1)t y,\psi_1) &+
    		b_2(t) y_\alpha p_\beta C^{dyn}((\alpha_2-\mu_2)t y,\psi_1)
    		+\\+
    		c_1(t) p_\alpha p_\beta C^{dyn}((\alpha_1-\mu_1)t y,\psi_1) &+
    		c_2(t) p_\alpha p_\beta C^{dyn}((\alpha_2-\mu_2)t y,\psi_1)
    		\big]\psi_1 \psi_2
    		\end{aligned}
	\end{gathered}
\end{equation}
with $a_i(t),b_i(t),c_i(t)$ of the form \eqref{diff_eq}. Demanding expressions \eqref{D0_ABC} to compensate  $\Delta R$ (\ref{D_R}) it is not hard to find the following conditions on the parameters:
\begin{equation}\label{bound_cond}
	\frac{b_1(0)}{\rho_1} + \frac{b_2(0)}{\rho_2} = -1; \qquad \frac{a_1(0)+c_1(0)}{\rho_1} + \frac{a_2(0)+c_2(0)}{\rho_2} = 1\,,\qquad \rho_i:=\alpha_i-\mu_i\,.
\end{equation}

\subsection{Family of shifts}

Using the explicit form of the polynomials in \eqref{dW_int} one can see that the condition \eqref{diff_eq} imposes the following restrictions on the homotopy parameters:
\begin{equation}\label{solution}
	\mu_1 = \nu_1 = \alpha_1-1;
	\quad
	\mu_2 = \nu_2 = \alpha_2+1;
	\quad
	\forall \alpha_1, \alpha_2\,.
\end{equation}
Then $\delta W_1$ acquires the form
\begin{equation}\label{dW_found}
	\begin{gathered}
		\delta W_1 =
		- \frac{\lambda}{4i} \int_0^1 dt h_0{}^{\alpha \beta} \times
		\\
		\begin{aligned}
    		\times \Big[
    		\frac{1}{2} \alpha_1 (1-t^2) y_\alpha y_\beta C^{dyn}(t y,\psi_1) &+
    		\frac{1}{2} \alpha_2 (1-t^2) y_\alpha y_\beta C^{dyn}(-t y,\psi_1)
    		-\\
    		-\frac{1}{2} (1-t^2) y_\alpha p_\beta C^{dyn}(t y,\psi_1) &+
    		\frac{1}{2} (1-t^2) y_\alpha p_\beta C^{dyn}(-t y,\psi_1)
    		+\\+
    		\frac{1}{2} (1-\alpha_1) (1-t^2) p_\alpha p_\beta C^{dyn}(t y\psi_1) &
    		-\frac{1}{2} (1+\alpha_2) (1-t^2) p_\alpha p_\beta C^{dyn}(-t y,\psi_1)
    		\Big]\psi_1 \psi_2,
		\end{aligned}
	\end{gathered}
\end{equation}
It is easy to see that it obeys the boundary conditions \eqref{bound_cond}. Note that the leftover freedom is in the $y$-shift parameters $\alpha_{1,2}$ in the homotopy operators.

Now it is interesting to compare this family of shifts $\delta W_1$ with  the shift $\delta \omega^{top}$ \eqref{dw_old} found in \cite{Vasiliev:1992ix}. To this end, we split $C^{dyn}(y,\psi_1)$ into  even and odd parts
\begin{equation}\label{EO}
C^{dyn}(y,\psi_1)=C^{dyn}_{+}(y,\psi_1)+C^{dyn}_{-}(y,\psi_1)\, ,
\end{equation}
{where}
\begin{equation}
C^{dyn}_+(y,\psi_1)=\frac{C^{dyn}(y,\psi_1)+C^{dyn}(-y,\psi_1)}{2}\, ,\;\; C^{dyn}_-(y,\psi_1)=\frac{C^{dyn}(y,\psi_1)-C^{dyn}(-y,\psi_1)}{2}.
\end{equation}
Here $C^{dyn}_+(y,\psi_1)$ and $C^{dyn}_-(y,\psi_1)$ describe bosons and fermions, respectively. In these terms, the difference between \eqref{dW_found} and \eqref{dw_old} takes form
\begin{equation}\label{coh}
    \begin{gathered}
        \delta W_1 - \delta \omega^{top} \psi_2 = - \frac{\lambda}{8i} \int_0^1 dt\, (1-t^2)  h_0^{\alpha \beta}
        \cdot
        \\
        \begin{aligned}
            \cdot
            \Big[
            \big( \alpha_1 - \frac{1}{2} \big) (y_{\alpha}y_{\beta} - p_{\alpha} p_{\beta}) C^{dyn}(t y,\psi_1) &+
            \big( \alpha_2 + \frac{1}{2} \big) (y_{\alpha}y_{\beta} - p_{\alpha} p_{\beta}) C^{dyn}(-t y,\psi_1)
            - \\ -
            (y_{\alpha}y_{\beta} + p_{\alpha} p_{\beta}) C_{+}^{dyn}(t y,\psi_1) &+
            2 y_{\alpha} p_{\beta} C_{-}^{dyn}(t y,\psi_1)
            \Big] \psi_1 \psi_2\,.
        \end{aligned}
    \end{gathered}
\end{equation}
On the {\it r.h.s.} one finds the first two terms to be $D_0$-exact and  the other two terms to be $D_0$-closed but non-exact. Note that here the parity of $C^{dyn}_{\pm}$ plays an important role. Indeed, by \eqref{ABC}, \eqref{D0_ABC}
\begin{subequations}\label{D0_coh}
\begin{align}
	D_0 \int_0^1 dt (1-t^2)  h_0^{\alpha \beta} (y_{\alpha}y_{\beta} + p_{\alpha} p_{\beta}) C_{+}^{dyn}(t y,\psi_1) \psi_1 \psi_2 \propto
	\left.H^{\alpha \beta} y_{\alpha} p_{\beta} C_{+}^{dyn}(\xi,\psi_1) \psi_2 \right|_{\xi=0}&=0\,,
	\\
	D_0 \int_0^1 dt (1-t^2)  h_0^{\alpha \beta} y_{\alpha} p_{\beta} C_{-}^{dyn}(t y,\psi_1) \psi_1 \psi_2
	\propto
	\left.H^{\alpha \beta}\left(y_{\alpha} y_{\beta}+p_{\alpha} p_{\beta}\right) C_{-}^{dyn}(\xi,\psi_1) \psi_2 \right|_{\xi=0}&=0\,.
\end{align}
\end{subequations}
Hence, the last two terms in the (\ref{coh}) belong to non-trivial $D_0$-cohomology.

Thus, as anticipated, the difference $(\delta W_1 - \delta \omega^{top} \psi_2)$ is $D_0$-closed, which implies that equation \eqref{Goal} is solved. However, the old shift $\delta \omega^{top}$ cannot be reproduced within the shifted homotopy  setup of this paper: free parameters $\alpha_1, \alpha_2$ only affect the gauge transformation, while $(\delta W_1 - \delta \omega^{top} \psi_2)$  belongs to a non-zero  $D_0$-cohomology class. It would be interesting to find an extension of the shifted homotopy approach rich enough to reproduce the $D_0$-cohomological terms as well and, in particular, the shift of variables of \cite{Vasiliev:1992ix}.

\subsection{Definite parity solutions}
Consider once again the equation \eqref{dz_W}. Schematically, it can be represented as \eqref{terms}. Since each term in the brackets on the {\it r.h.s.} is  $\dr_z$-closed, one can use different homotopy operators for each of them. However, there is an additional freedom that is not considered so far.
One can decompose $C^{dyn}$ into even and odd parts \eqref{EO}. Then, using that
$\dr_z C^{dyn}_\pm(y,\psi_1)=0$, $\dr_z h_0(y)=0$, $\dr_z \gamma=0$, $h_q\gamma=0$ and formula \eqref{resId}, one
can show that the following expressions
\begin{equation}
\frac{\lambda}{4 i}h_{0} * C_{\pm}^{dyn} *  \big(\Delta_{p+t}- \Delta_{p}\big) \gamma\psi_1 \psi_2 k\,,\qquad
\frac{\lambda}{4 i}C_{\pm}^{dyn} * \tilde{h}_0 * \big(\Delta_{p-2t}-\Delta_{p-t}\big) \gamma  \psi_1 \psi_2 k
\end{equation}
are  $\dr_z$-closed as well,  which allows one to apply different homotopy operators to each of them.


As in Section \ref{Shifted_homotopy_setup}, we apply non-conventional homotopy only to some of the terms on the {\it r.h.s.} of \eqref{dz_W}, since the other resulting from the conventional homotopy happen to be covariantly constant.  Since each term of
\begin{multline}
\frac{\lambda}{4 i}h_{0} * C^{dyn}_+ *  \big(\Delta_{p+t}- \Delta_{p}\big) \gamma\psi_1 \psi_2 k+\frac{\lambda}{4 i}C^{dyn}_+ * \tilde{h}_0 * \big(\Delta_{p-2t}-\Delta_{p-t}\big) \gamma  \psi_1 \psi_2 k+\\
+\frac{\lambda}{4 i}h_{0} * C^{dyn}_- *  \big(\Delta_{p+t}- \Delta_{p}\big) \gamma\psi_1 \psi_2 k+\frac{\lambda}{4 i}C^{dyn}_- * \tilde{h}_0 * \big(\Delta_{p-2t}-\Delta_{p-t}\big) \gamma  \psi_1 \psi_2 k
\end{multline}
is $\dr_z$-closed, this allows us to apply $\Delta_{\mu^+_1p+\nu^+_1 t+\alpha^+_1 y}$,$\Delta_{\mu^+_2p+\nu^+_2 t+\alpha^+_2 y}$, $\Delta_{\mu^-_1p+\nu^-_1 t+\alpha^-_1 y}$, $\Delta_{\mu^-_2p+\nu^-_2 t+\alpha^-_2 y}$  to the first, second, third and fourth term, respectively. The difference between $W_1$ obtained  this way and \eqref{W_1^0} is
\begin{equation}
W_1-W_1^0=\delta W_{1}^++\delta W_{1}^{-},
\end{equation}
where $\delta W_1^\pm$ have the form of \eqref{dW_int} but with respective $\pm$-superscripts.
Note that $\Delta R^{top}$ also decomposes into bosonic and fermionic parts as
\begin{equation}\label{R+}
    \Delta R_+^{top}\left(y ; \psi_{1}\right)=-\left.\frac{\lambda^2}{16i} h_{0}{ }^{\alpha}{}_{\gamma} \wedge h_{0}{}^{\gamma \beta}\left(y_\alpha y_\beta + p_\alpha p_\beta \right) C_+^{dyn}\left(\xi ; \psi_{1}\right)\right|_{\xi=0}\,,
\end{equation}
\begin{equation}\label{R-}
    \Delta R_-^{top}\left(y ; \psi_{1}\right)=-\left.\frac{\lambda^2}{16i} h_{0}{ }^{\alpha}{}_{\gamma} \wedge h_{0}{}^{\gamma \beta}\left(y_\alpha p_\beta + p_\alpha y_\beta \right) C_-^{dyn}\left(\xi ; \psi_{1}\right)\right|_{\xi=0}.
\end{equation}
Analogously to the previous section, taking into account the form of expressions \eqref{dW_int} and \eqref{R+}, \eqref{R-}, one should choose $(\rho^\pm_i)^2=1$, where $\rho^\pm_i=\alpha_i^\pm - \mu_i^\pm$ (cf. \eqref{ABC}, \eqref{diff_eq}).  For example, consider $\delta W_1^+$ at $\rho^+_1=1$ and $\rho^+_2=-1$
\begin{equation}\label{dW+}
    \begin{gathered}
        \delta W^{+}_1=-\frac{\lambda}{4i}\int_0^1 d t\, (1-t)\, h_0^{\alpha \beta} \times\\
        \times \Big\{\Big[(\alpha^+_1+\alpha^+_2)\frac{1-t}{2}-\alpha^+_1(\nu^+_1-\alpha^+_1)t+\alpha^+_2(\nu^+_2-\alpha^+_2)t \Big]y_\alpha y_\beta C^{dyn}_{+}(t y,\psi_1)+\\
        +\Big[(\alpha^+_1-1)\Big(\frac{1-t}{2}-(\nu^+_1-\alpha^+_1)t\Big)-\alpha^+_1\Big(\frac{1+3t}{2} t+(\nu^+_1-\alpha^+_1)t\Big)\Big]y_\alpha p_\beta C^{dyn}_{+}(t y,\psi_1)+\\
        +\Big[(\alpha^+_2+1)\Big(\frac{1-t}{2}+(\nu^+_2-\alpha^+_2)t\Big)-\alpha^+_2\Big(\frac{1+3t}{2}-(\nu^+_2-\alpha^+_2)t\Big)\Big]y_\alpha p_\beta C^{dyn}_{+}(t y,\psi_1)-\\
        -\Big[(\alpha^+_1-1)\Big(\frac{1+3t}{2}+(\nu^+_1-\alpha^+_1)t\Big)+(\alpha^+_2+1)\Big(\frac{1+3t}{2}-(\nu^+_2-\alpha^+_2)t\Big)\Big]p_\alpha p_\beta C^{dyn}_{+}(t y,\psi_1)\Big\}\psi_1 \psi_2.
    \end{gathered}
\end{equation}
Here it is used that $C^{dyn}_+(y,\psi_1)$ is  even  in $y$. We wish \eqref{dW+} to solve equation $D_0(\delta W_1^+)=\Delta R_+^{top}\psi_2$. Using the rules \eqref{diff_eq} and \eqref{bound_cond}, one finds the same constraints on the homotopy parameters as \eqref{solution}:
\begin{equation}
\mu^+_1 = \nu^+_1 = \alpha^+_1-1,\quad \mu^+_2 = \nu^+_2 = \alpha^+_2+1\,.
\end{equation}
For $\rho ^{+}_{1}=\rho ^{+}_{2}$ there are no solutions for $\mu ^{+}_{1,2},\nu ^{+}_{1,2},\alpha ^{+}_{1,2}$ compatible with the requirement $D_{0}(\delta W_{1}^{+})=\Delta R_{+}^{top} \psi_2$, while for $\rho ^{+}_{1}=-1, \rho ^{+}_{2}=1$ the solution is
\begin{equation}
	\mu^+_1 = \alpha^+_1+1,
	\mu^+_2 = \alpha^+_2-1,
	\quad
	\nu^+_1 = \alpha^+_1 - \frac{1}{1+2 \alpha^+_1},
	\nu^+_2 = \alpha^+_2 + \frac{1}{1-2 \alpha^+_2},
	\quad
	\alpha^+_1 \neq -\frac{1}{2},
	\alpha^+_2 \neq \frac{1}{2}\,.
\end{equation}

Analogously, one finds homotopy parameters for $\delta W_1^-$ providing $D_0(\delta W_1^-)=\Delta R_-^{top}\psi_2$. Namely, using \eqref{diff_eq} and \eqref{bound_cond} one finds the following constraints on the fermionic homotopy parameters: at $\rho_1^- = 1, \rho_2^- = -1$
\begin{equation}
\mu^-_1 = \nu^-_1 = \alpha^-_1-1,\quad \mu^-_2 = \nu^-_2 = \alpha^-_2+1
\end{equation}
and at $\rho_1^- = -1, \rho_2^- = 1$
\begin{equation}
	\mu^-_1 = \alpha^-_1+1,\;
	\mu^-_2 = \alpha^-_2-1,\;
		\nu^-_1 = \alpha^-_1 + \frac{1}{1+2 \alpha^-_1},\;
	\nu^-_2 = \alpha^-_2 - \frac{1}{1-2 \alpha^-_2},\;
	\alpha^-_1 \neq -\frac{1}{2}, \;
	\alpha^-_2 \neq \frac{1}{2}\,.
\end{equation}
Again, one can check that being  $D_0$-closed the expression $\left( \delta W_1^++\delta W_1^--\delta \omega^{top}\psi_2 \right)$ is not $D_0$-exact for all allowed $\alpha^\pm_i$.

\section{$D_0$-cohomology from massive terms}

In the previous section we have found two types of $D_0$-closed but not $D_0$-exact expressions \eqref{coh}, \eqref{D0_coh}.
Let us set
\begin{subequations}\label{D0_closed}
\begin{align}
	G^{+} &= \frac{\lambda}{8i} \int_0^1 dt (1-t^2)  h_0^{\alpha \beta} (y_{\alpha}y_{\beta} + p_{\alpha} p_{\beta}) C_{+}^{dyn}(t y,\psi_1) \psi_1 \psi_2,
	\\
	G^{-} &= -\frac{\lambda}{4i} \int_0^1 dt (1-t^2)  h_0^{\alpha \beta} y_{\alpha} p_{\beta} C_{-}^{dyn}(t y,\psi_1) \psi_1 \psi_2\,.
\end{align}
\end{subequations}
Since $G^\pm$ are $D_0$-closed, they can be added to the {\it r.h.s.} of the covariant constancy equation  for $C(y)$ without violating  its consistency. It can be shown that these terms reproduce the linearized mass corrections in the $AdS_3$ HS theory described in \cite{Barabanshchikov:1996mc} (for more information on description massive higher spin particles in AdS see also \cite{Zinoviev:2001dt,Zinoviev:2008ze,Khabarov:2022hbv}). The latter have different forms for bosons and fermions:
\begin{subequations}\label{mass_old}
\begin{align}
&D C^{boson}_{\alpha(n)}=\frac{\psi_1}{2i} \left[\left(1-\frac{\nu(\nu - 2)}{(n+1)(n+3)}\right) h^{\beta \gamma} C_{\beta \gamma \alpha(n)}-\lambda^{2} n(n-1) h_{\alpha \alpha} C_{\alpha(n-2)}\right],
\\
&\begin{aligned}
D C^{fermion}_{\alpha(n)}= \frac{\psi_1}{2i} \left(1-\frac{\nu^{2}}{(n+2)^{2}}\right) h^{\beta \gamma} C_{\beta \gamma \alpha(n)} - \nu &\frac{\lambda \psi_1}{n+2} h_{\alpha}{}^{\beta} C_{\beta \alpha(n-1)}
- \\ -
&\frac{\psi_1}{2i} \lambda^{2} n(n-1) h_{\alpha \alpha} C_{\alpha(n-2)}.
\end{aligned}
\end{align}
\end{subequations}
Here $\nu$ is a parameter of the deformed Weyl algebra $\left[\hat{y}_{\alpha}, \hat{y}_{\beta}\right]=2 i \epsilon_{\alpha \beta}(1+\nu k)$ \cite{Vasiliev:1989re} related to the mass.

This can be seen by rewriting $G^\pm$ \eqref{D0_closed} in the component form analogous to \eqref{C_comp}
\begin{subequations}
\begin{align}
    &G^{+}_{\alpha(n=2k)} = - \frac{\psi_1}{4i} \left[\frac{1}{(n+1)(n+3)} h_0^{\beta \gamma} C^{dyn}_{\beta \gamma \alpha(n)} - \lambda^2 n(n-1) \frac{1}{(n-1)(n+1)} h_0{}_{\alpha \alpha} C^{dyn}_{\alpha(n-2)}
    \right]\,,
\\
    &G^{-}_{\alpha(n=2k-1)} = - \frac{1}{2} \frac{\lambda \psi_1}{n+2}  h_{0 \alpha}{}^{\beta} C^{dyn}_{\beta \alpha(n-1)}\,,
\end{align}
\end{subequations}
From \eqref{D0_closed} it is also obvious that $G^{+}_{\alpha(n=2k-1)}=G^{-}_{\alpha(n=2k)}=0$.

So, one can deform the unfolded equation \eqref{DC_unfolded} as
\begin{equation}
    D_0 C = a G^{+} + b G^{-}\,.
\end{equation}
The resulting deformed equations differ for even and odd $n$:
\begin{subequations}\label{mass_new}
\begin{align}
    &\begin{aligned}
    D C_{\alpha(n=2k)} = \frac{\psi_1}{2i} \bigg[&\left(1-\frac{a/2}{(n+1)(n+3)}\right) h_0^{\beta \gamma} C^{dyn}_{\beta \gamma \alpha(n)}
    -\\-
    \lambda^{2} n(n-1) &\left(1-\frac{a/2}{(n-1)(n+1)}\right) h_{0 \alpha \alpha} C^{dyn}_{\alpha(n-2)}\bigg],
    \end{aligned}
\\
    &D C_{\alpha(n=2k-1)} = \frac{\psi_1}{2i}  h_0^{\beta \gamma} C^{dyn}_{\beta \gamma \alpha(n)} + \frac{b}{2} \frac{\lambda \psi_1}{n+2} h_{0 \alpha}{}^{\beta} C^{dyn}_{\beta \alpha(n-1)} - \frac{\psi_1}{2i} \lambda^{2} n(n-1) h_{0 \alpha \alpha} C^{dyn}_{\alpha(n-2)}.
\end{align}
\end{subequations}

It can be seen that deformed equations \eqref{mass_new} for $C_{\alpha(n)}$ with even and odd $n$ reproduce respectively the linearized massive equations for bosons and fermions \eqref{mass_old} with
\begin{equation}
    a = 2 \nu (\nu-2), \quad b = -2 \nu.
\end{equation}

\section{Vertex $\Upsilon(\omega, \omega, C)$}

In the higher orders of the perturbative expansion, one has to evaluate the {\it r.h.s.} of the field equations. Here we show how the  vertex $\Upsilon(\omega, \omega, C)$ can be obtained within the shifted homotopy approach for arbitrary (not necessarily $AdS_3$)  $\omega_0(y)$ and $h_0(y)$ verifying \eqref{PV_1.0}.

So, our goal is to find a solution $W_1 = \omega_1 + W_1^q$ of equation \eqref{dz_W}
\begin{equation*}
	\dr_z W_1 = - \frac{1}{4i} \left[ W_0 * C * (\Delta_{p}-\Delta_{p+t})\gamma - C * \tilde{W}_0 * (\Delta_{p-2t}-\Delta_{p-t})\gamma \right]k
\end{equation*}
which provides $D_0 \omega_1 = 0$. Therefore, we have to choose the shift $q$ so that $D_0 \omega_1 = D_0 (-W_1^q) = 0$. Since $W_1^q = W_1^0 + \delta W$, where $W_1^0$ and $\delta W$ are expressed as \eqref{W_1^0} and \eqref{dW_ddg}, respectively, the following formulas valid up to higher orders have to be used to calculate $D_0(\delta W)$:
\begin{equation}
	\dr h_0(y) = - \omega_0(y) * h_0(y) - h_0(y) * \omega_0(y)\,,
\end{equation}
\begin{equation}
	\dr C^{dyn}(y) = - \omega_0(y) * C^{dyn}(y) + C^{dyn}(y) * \tilde{\omega}_0 - \psi_1 h_0(y) * C^{dyn}(y) - \psi_1 C^{dyn}(y) * \tilde{h}_0\,.
\end{equation}

It can be seen that the vertex $D_{0}\left(\omega_1 \right)$  consists of three parts: the known $\Delta R$ \eqref{D_R}, the vertex $\Upsilon_{\omega h}$, proportional to $\left(\omega_{0}, h_{0}\right)$, and  $\Upsilon_{h h}$, proportional to $ \left(h_{0}, h_{0}\right)$
\begin{equation}
	D_0(\omega_1)=\Delta R + \Upsilon_{\omega h} + \Upsilon_{h h}.
\end{equation}

For the $AdS_3$-connection, one can check that the $(\omega_0, h_0)$-vertex is identically  zero $\Upsilon_{\omega h} \equiv 0$ regardless of the homotopy parameters\footnote{This is not incidental, being a consequence of the Lorentz covariance of the nonlinear equations of \cite{Prokushkin:1998bq} as well as of the Lorentz covariance of the homotopy procedure.}.

Direct calculation using the star-exchange formulas gives (here $C$ reads as $C^{dyn} \psi_2$):
\begin{equation}\label{U_hh}
	\begin{aligned}
		\Upsilon_{h h} = \frac{1}{4i}
		\Big[
		&+h_0 * h_0 * C *
		h_{(\mu_1-\alpha_1+1)p+(\nu_1-\alpha_1+1)t_1+(\mu_1-\alpha_1+1)t_2+\alpha_1 y} \Delta_{p+t_1+t_2} \Delta_{p+t_2} \gamma
		\\
		&+h_0 * h_0 * C *
		h_{(\mu_1-\alpha_1+1)p+(\nu_1-\alpha_1+1)t_2+\alpha_1 y} \Delta_{p+t_2} \Delta_{p} \gamma
		\\
		&+h_0 * C * \tilde{h}_0 *
		h_{(\mu_1-\alpha_1+1)p+(\nu_1-\alpha_1+1)t_1-(\mu_1-\alpha_1+1)t_2+\alpha_1 y} \Delta_{p+t_1-t_2} \Delta_{p-t_2} \gamma
		\\
		&-h_0 * C * \tilde{h}_0 *
		h_{(\mu_1-\alpha_1+1)p+(\nu_1-\alpha_1+1)t_1-2t_2+\alpha_1 y} \Delta_{p+t_1-2t_2} \Delta_{p-2t_2} \gamma
		\\
		&-h_0 * C * \tilde{h}_0 *
		h_{(\mu_2-\alpha_2+1)p+(\mu_2-\alpha_2+1)t_1-(\nu_2-\alpha_2+1)t_2+\alpha_2 y} \Delta_{p+t_1-t_2} \Delta_{p+t_1-2t_2} \gamma
		\\
		&+h_0 * C * \tilde{h}_0 *
		h_{(\mu_2-\alpha_2+1)p-(\nu_2-\alpha_2+1)t_2+\alpha_2 y} \Delta_{p-t_2} \Delta_{p-2t_2} \gamma
		\\
		&-C * \tilde{h}_0 * \tilde{h}_0 *
		h_{(\mu_2-\alpha_2+1)p-(\mu_2-\alpha_2+1)t_1-(\nu_2-\alpha_2+1)t_2+\alpha_2 y} \Delta_{p-t_1-t_2} \Delta_{p-t_1-2t_2} \gamma
		\\
		&-C * \tilde{h}_0 * \tilde{h}_0 *
		h_{(\mu_2-\alpha_2+1)p-(\nu_2-\alpha_2+1)t_1-2t_2+\alpha_2 y} \Delta_{p-t_1-2t_2} \Delta_{p-2t_1-2t_2} \gamma
		\Big].
	\end{aligned}
\end{equation}

By substituting the $AdS_3$-connection, one can find restrictions on the homotopy parameters under which $\Delta R + \Upsilon_{hh} = 0$, which coincide with those   found in Section \ref{section7}:
\begin{equation}
	\mu_1 = \nu_1 = \alpha_1-1;
	\quad
	\mu_2 = \nu_2 = \alpha_2+1;
	\quad
	\forall \alpha_1, \alpha_2\,.
\end{equation}

\section{Conclusion}
In this paper a $3d$ nonlinear system of HS equations  has been analyzed in the first order in the zero-forms using the shifted homotopy technique.
The $3d$ HS theory differs from the $4d$ theory of \cite{Vasiliev:1992av} in that respect that in the latter the conventional (zero-shift) homotopy leads directly to the proper form of the free HS equations  while in the former it leads to an entanglement between dynamical and topological fields.
The main result of this paper, anticipated to  simplify the analysis of the higher-order corrections to nonlinear HS field equations, consists of finding a shifted homotopy procedure that eliminates this entanglement.

Interestingly enough, the resulting field redefinition differs from that found originally in \cite{Vasiliev:1992ix} by the direct analysis without using the shifted homotopy technique.
This difference is shown to be cohomological in nature being related to the  deformed oscillator algebra underlying the massive deformation of the matter field equations in $3d$ HS theory. It raises the question whether there exists an extension of the shifted homotopy approach rich enough to reproduce the field redefinition of \cite{Vasiliev:1992ix} as well.

To summarize, we reached a coherent picture, which will allow us in the future to perform higher-order analysis of $3d$ HS theory with the proper disentanglement of the dynamical and topological fields.
The applied method of shifted homotopy is very convenient and greatly simplifies the analysis.
In particular, it will be interesting  to see how the problem of disentangling of the dynamical and topological fields can be resolved at higher orders.

\section*{Acknowledgement}

This research was
supported by the Russian Science Foundation grant 18-12-00507.




\end{document}